\documentclass[prl,preprint,showpacs,preprintnumbers,amsmath,amssymb]{revtex4}
\usepackage{graphicx}
\usepackage{dcolumn}
\usepackage{bm}
\usepackage{mathrsfs}

\begin{document}

\title{Large-amplitude chirped coherent phonons in tellurium
mediated by ultrafast photoexcited carrier diffusion}

\author{N. Kamaraju,~Sunil Kumar, M. Anija and~A. K. Sood \footnote{Electronic mail:~asood@physics.iisc.ernet.in}}

\affiliation{Department of Physics and Center for Ultrafast
Laser Applications (CULA), Indian Institute of
Science,~Bangalore - 560 012, India }

\date{\today}

\begin{abstract}
We report femtosecond time-resolved reflectivity
measurements of coherent phonons in tellurium performed
over a wide range of temperatures (3K to 296K) and pump
laser intensities. A totally symmetric A$_{1}$ coherent
phonon at 3.6 THz responsible for the oscillations in the
reflectivity data is observed to be strongly positively
chirped (i.e, phonon time period decreases at longer
pump-probe delay times) with increasing photoexcited
carrier density, more so at lower temperatures. We show for
the first time that the temperature dependence of the
coherent phonon frequency is anomalous (i.e, increasing
with increasing temperature) at high photoexcited carrier
density due to electron-phonon interaction. At the highest
photoexcited carrier density of $\sim$ 1.4 $\times$
10$^{21}$cm$^{-3}$ and the sample temperature of 3K, the
lattice displacement of the coherent phonon mode is
estimated to be as high as $\sim$ 0.24 \AA. Numerical
simulations based on coupled effects of optical absorption
and carrier diffusion reveal that the diffusion of carriers
dominates the non-oscillatory electronic part of the
time-resolved reflectivity. Finally, using the pump-probe
experiments at low carrier density of 6 $\times$ 10$^{18}$
cm$^{-3}$, we separate the phonon anharmonicity to obtain
the electron-phonon coupling contribution to the phonon
frequency and linewidth.
\end{abstract}


\maketitle
\subsection{I. INTRODUCTION}
The investigation of ultrafast carrier and phonon dynamics in
semiconductors is of special interest as their applications in
information technology depends on the understanding of fundamental
processes like momentum and energy relaxation, as well as
mechanisms such as carrier-carrier scattering, inter-valley and
intra-valley scattering, optical phonon scattering, and carrier
diffusion \cite{Te_A_Othonos_JAP_Review_1998}. When a
semiconductor is irradiated with an ultrashort laser pulse with
photon energy much higher than the band gap, it excites electrons
from valence band to the conduction band with excess of energy
above the bottom of the conduction band. The carrier distribution
immediately after the photoexcitation is of non-Fermi-Dirac type
(nonthermal) \cite{Te_Knox_PRL} where the temperature of the
excited carriers is still not defined. Subsequently the
carrier-carrier interactions lead to a formation of Fermi-Dirac
distribution with a definite electron temperature (also called as
thermalization of carriers) from the initial nonthermal state and
this process depends on the excited carrier density. At moderate
carrier density of $\sim$ 10$^{18}$ cm$^{-3}$, this
carrier-carrier interaction time is $\sim$ 200 fs in GaAs
\cite{Te_Knox_PRL, Te_Oudar_PRL} which can reduce to a few fs at
higher photoexcited carrier densities. Following this, the hot
carriers diffuse in the material as well as emit phonons. The
magnitude of cooling time again depends on the number of excited
carriers and the initial temperature of the distribution.
Typically, carrier cooling times are of the order of a few
picoseconds in semiconductors \cite{Te_J_Shaw}. There is, finally,
the electron-hole recombination which happens on the time scale of
several picoseconds to nanoseconds.

    In recent years, the femtosecond pulses have been used
to generate and probe coherent phonons in solids \cite{Te_Nelson,
Te_DECP_PRB_1992, Te_Cheng_APL, Te_Kuznetsov, Te_Garret_PRL_1996}.
The excitation mechanism of coherent phonons was thought to be
impulsive stimulated Raman scattering (ISRS) \cite{Te_Nelson} in
transparent materials whereas displacive excitation of coherent
phonons (DECP) is the dominant mechanism in opaque samples
\cite{Te_DECP_PRB_1992, Te_Cheng_APL, Te_Kuznetsov}. Later, it was
shown that DECP is a special case of ISRS when excited resonantly
\cite{Te_Garret_PRL_1996}. In a resonant ISRS/DECP process, the
photo-excitation of electrons changes the equilibrium positions of
the atoms; the atoms then oscillate around their new equilibrium
positions. In absorbing samples like Te, the Dember field arising
from different diffusion coefficients of photo-excited electrons
and holes can also contribute to the generation of coherent
phonons \cite{Te_Dember, Te_Kurz_PRL, Te_Kurz_PRB}.

    Till now, there have been only a very few ultrafast
coherent phonon studies at photoexcited carrier densities
(PCD) as high as 10$^{20}$ cm$^{-3}$ \cite{Te_Hunsche,
Te_Hunsche_APA, Te_Tangney_PRB_2002, Te_DeCamp,
Te_M_Hase_PRL_2002, Te_Misochko_PRL_2004, Te_Murray}. A red
shift of frequency of the totally symmetric $A_{1}$ phonon
was observed in a narrow band gap semiconductor, tellurium,
as the femtosecond laser injected carrier density was
increased to 5 $\times$ 10$^{21}$ cm$^{-3}$ at room
temperature in a degenerate pump probe reflectivity
experiments \cite{Te_Hunsche, Te_Hunsche_APA}. This was
attributed to the bond weakening by the carriers
\cite{Te_Stampfi_PRB_electron_Softening}. In addition, the
phonon time period was found to decrease with increase in
the delay time between pump and probe pulses, leading to
asymmetric line-shape in frequency domain
\cite{Te_Hunsche}. This linear sweep in the frequency with
the pump-probe time delay can be termed as phonon chirping
which originates from the rapid change of photoexcited
carrier density in the penetration depth of the sample due
to carrier diffusion and its interaction with the coherent
phonons. Similar phonon chirping has been seen in
degenerate pump probe experiments on bismuth
\cite{Te_M_Hase_PRL_2002, Te_Misochko_PRL_2004} with PCD of
$\geq$ 3 $\times$ 10$^{21}$cm$^{-3}$. Two pump pulse
experiments \cite{Te_Hunsche, Te_Murray} along with first
principal density functional calculations
\cite{Te_Tangney_PRB_2002, Te_Murray} have established that
the chirping arises from varying carrier density along the
probed length in the sample and not due to lattice
anharmonic effects.

    We note here that all the previous studies on tellurium have
been performed at room temperature. Here, we present
femtosecond time-resolved reflectivity measurements of
coherent phonon dynamics in tellurium performed over a wide
range of temperatures (3K to 296K) and carrier density
levels. The non-oscillatory part of the data associated
with the strength of photoexcited carrier density is found
to be dominated by diffusion of the carriers as revealed by
our numerical simulations. The oscillatory part shows
highly positively chirped A$_1$ phonons excited at
photoexcited carrier density of $\sim$ 1.4 $\times$
10$^{21}$ cm$^{-3}$ and the chirping is found to be largest
at lower temperatures where the carrier diffusion time and
the strength of photoexcited carrier density were found to
be maximum. Apart from the phonon chirping, the associated
lattice displacement is estimated to be as high as $\sim$
0.24 \AA~ at 3K. The other important result is the
observation of increased softening of $A_1$ mode with the
pump fluence at 3K (-0.16 THz/mJ-cm$^{-2}$) as compared to
296K (-0.08 THz/mJ-cm$^{-2}$) consistent with the increased
photoexcited carriers at lower temperatures. Using
time-resolved reflectivity experiments performed at low
carrier density levels of $\sim$ 6 $\times$ 10$^{18}$
cm$^{-3}$ where phonon chirping is found to be absent, the
effects of phonon anharmonicity and high carrier density
are separated and analysed.

\section{II. SAMPLE DETAILS}
In our experiments, we have used a polycrystalline
tellurium crystal (6$\times$4$\times$2 mm$^3$) mounted on a
continuous helium flow optical cryostat. Tellurium is a
narrow band semiconductor with a band gap of $\sim$ 0.33 eV
\cite{Te_Tangney_PRB_2002}. It crystallizes into $P3_121$
structure with the point group $D_3^4$ and three atom basis
arranged around a screw axis parallel to the \emph{c} axis
of the hexagonal unit cell forming a helical structure. The
equilibrium lattice constants are \emph{a} =
4.456~\AA~(distance between two helices) and \emph{c} =
5.927~\AA. The interatomic spacing in a chain is
2.834~\AA~and the helical radius is \emph{u} = 1.173 ~\AA.
Group theory gives the allowed Raman modes
\cite{Te_pine_raman} as A$_1$ ($\sim$ 3.6 THz)+
E$^1_{TO}$($\sim$ 2.8 THz) + E$^2_{TO}$($\sim$ 4.2 THz).
The totally symmetric A$_1$ mode corresponds to symmetric
inter-chain dilation and compression normal to the chain
axis, involving bond length and bond angle distortions. In
this mode, \emph{a} and \emph{c} do not change whereas only
helical radius \emph{u} changes.

The linear absorption coefficient \cite{Te_Absorption_PR},
$\alpha$, of Te at 1.57 eV is 2 $\times 10^5$ cm$^{-1}$
which gives the penetration depth $\xi$ ($\sim 1/\alpha$)
of pump light to be $\sim$ 50 nm. At maximum pump fluence
of 3.3 mJ/cm$^2$, the excited carrier density is $N_0 =
F\alpha(1-R)/E_p$ is $\sim$ 1.4 $\times 10^{21}$ cm$^{-3}$
which is 0.8 $\%$ of the total density of electrons in the
valence band. Here, $F$ is the pump fluence in J/cm$^2$,
$R$ is the reflectivity (0.46 at 1.57 eV) of tellurium and
E$_p$ is the photon energy in J.

\section{III. EXPERIMENTAL DETAILS}
Degenerate femtosecond pump-probe experiments in the reflection
geometry were carried out using femtosecond pulses derived from
Ti:Sapphire amplifier producing 50 fs pulses with the central
photon energy of 1.57 eV at a repetition rate of 1 kHz. The pump
beam was modulated at 393 Hz with a mechanical chopper and the
reflected probe intensity was recorded with a Si-PIN diode and a
lock-in amplifier. The pulse width at the sample point was 65~fs
(FWHM) measured using a thin beta-barium borate (BBO) crystal. The
spot sizes (half width at 1/e maximum) of the pump and the probe
beams were kept at 600 $\mu$m and 400 $\mu$m, respectively at the
overlap of the two beams at the sample. The difference between the
probe intensity reflected from the sample and that of a reference
beam was recorded as a function of delay between the pump and
probe pulses. Both the pump and probe beams were kept close to the
normal incidence with their polarizations perpendicular to each
other. This detection scheme can detect only A$_{1}$ mode for a
single crystal. However, the non-symmetric E modes can be seen as
well for a polycrystalline sample with major crystal axes oriented
at an angle from the surface \cite{Te_Kurz_PRL,
Te_M_Hase_PRL_2002}. In our experiments, we observed that the
damage threshold is $\sim$ 4.5 mJ/cm$^2$ which is much smaller
than the value of 12 mJ/cm$^2$ reported earlier \cite{Te_Hunsche}.
The pump fluence was varied between 3.3 mJ/cm$^2$ and 0.3
mJ/cm$^2$ whereas the probe fluence was kept at 0.04 mJ/cm$^2$.
The time-resolved reflectivity was recorded at four temperatures -
3K, 80K, 160K and 296K. At the lowest pump fluence of 14
$\mu$J/cm$^2$ (probe fluence $\sim$ 0.8 $\mu$j/cm$^2$ ), the
transient reflectivity data was recorded as a function of
temperature from 296K to 3K using femtosecond pulses derived from
the oscillator (50fs, Tsunami, 76MHz, Spectra Physics Inc) in the
fast scan scheme \cite{Te_Shaker}. In this scheme the pump delay
line is modulated with a shaker at 65 Hz and the time-resolved
signal was detected with a fast A/D converter (AIXscan data
acquisition system, AMO GmbH, Germany Inc.) which helps in
capturing relative signal changes of $\sim 10^{-7}$ without using
a lock-in amplifier \cite{Te_Shaker}.

\section{IV. RESULTS AND DISCUSSION}
\begin{figure}[htb]
\centering\includegraphics[scale=0.85]{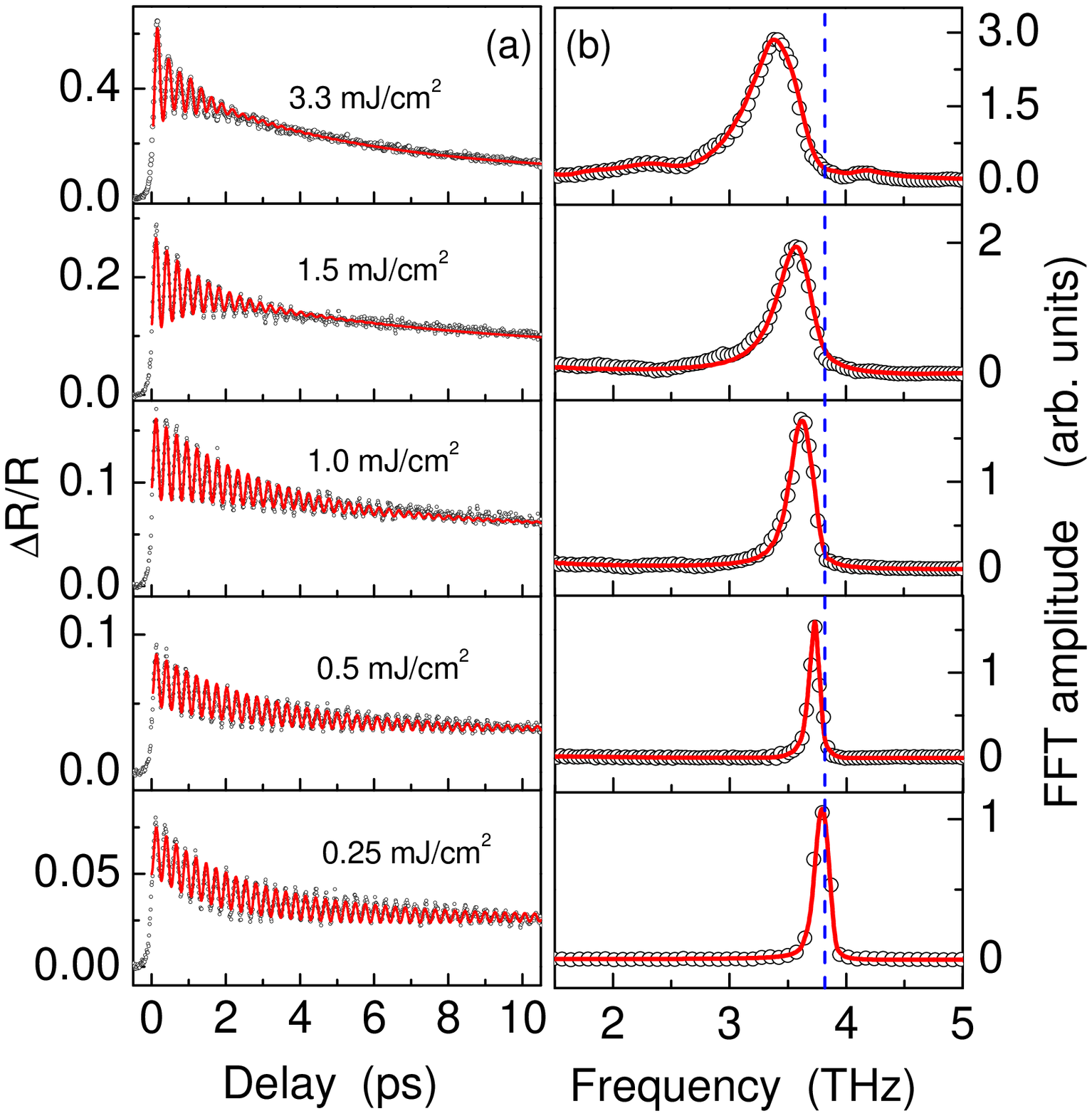}
\caption{}\label{Te_Fig1}
\end{figure}

The summary of results at 3K for various pump fluences is
presented in Fig.\ref{Te_Fig1} where the panel (a) is for
the time domain data and panel (b) shows the Fast Fourier
Transform (FFT) of the oscillatory part of the data. The
signal contains damped oscillations associated with the
generated coherent phonons overlaid on an exponentially
decaying background. This background can be assigned to
direct changes in electronic susceptibility by the
photoexcited carriers \cite{Te_Hunsche}. The vertical
dashed line in frequency domain is a guide to the eye to
appreciate the red shift of the peak frequency with
increased pump fluence. The time domain data at the lowest
pump fluence of 0.25 mJ/cm$^2$ can be fitted with the
following equation
\begin{equation}\label{Te_DR_by_R}
    \frac {\Delta R} {R} = y_0 + A_{el}~\mbox{exp}(-t/\tau_{el})+\displaystyle B
    ~\mbox{exp}(-t/\tau)~cos(2 \pi \nu t + \phi)
\end{equation}
where $\tau_{el}$ is electronic relaxation time; and
$A_{el}\equiv (\frac{\Delta R}{R})_{el}$  is the strength
of electronic relaxation contributions; $B\equiv
(\frac{\Delta R}{R})_{ph}$, $\tau$, $\nu$, and $\phi$
denote the amplitude, relaxation time, frequency, and
initial phase of the coherent phonon oscillations,
respectively. Here y$_0$ denotes the background arising
from long relaxation time constant ($>$1 ns) which may be
associated with slow electron-hole recombination time. At
higher pump intensities, the fit to the data using Eq.
\eqref{Te_DR_by_R} was found to be unsatisfactory as the
frequency rendering a good fit at short time delays
deviates considerably from the data at longer time delays.
This is demonstrated in Fig. \ref{Te_Fig2} (a) and (b) for
the time domain reflectivity data and FFT of its
oscillatory part at 3K and pump fluence of 3.3 mJ/cm$^2$
where thin grey line is the fit using Eq.
\eqref{Te_DR_by_R}. The experimental data is best fitted by
taking into account of phonon chirping. The transient
reflectivity data is modelled as
\begin{equation}\label{Te_DR_R_chirp}
    \frac {\Delta R} {R} = y_0 + A_{el}~\mbox{exp}(-t/\tau_{el})+\displaystyle\sum_{i=1,2,3} B_{i}
    ~\mbox{exp}(-t/\tau_i)~cos(2 \pi \nu_i t + \beta_i t^2 + \phi_i)
\end{equation}
where $\beta$ is the chirp parameter describing linear sweep in
the phonon frequency with the pump-probe delay time. Here the fit
is carried out with three modes (represented by subscript,
\emph{i} = 1, 2 and 3) guided by the FFT (Fig. \ref{Te_Fig2} (b)).
The coherent phonon modes observed at 2.45 THz, 3.25 THz and 4.13
THz are attributed to E$_{TO}^1$,~A$_{1}$ and E$_{TO}^2$,
respectively. It is found that the Eq. \eqref{Te_DR_R_chirp}
furnishes a very good fit (black line in Fig. \ref{Te_Fig2}(a) and
\ref{Te_Fig2}(b)) to the data. The chirping is significant for
only the A$_1$ mode (\emph{i} = 2) for which the obtained fitting
parameters are $\beta_2$ = 0.55 ps$^{-2}$, $\nu_2$ = 3.24 THz,
$\tau_2$ = 0.81 ps and B$_2$ =~0.22. The value of A$_1$ phonon
oscillation amplitude, B$_2$ is 34$\%$ of the maximum change in
reflectivity, $(\frac{\Delta R}{R})_{max}$( = 0.648). We note that
the non-symmetric E-modes were observed only at 3K using
pump-fluence of 3.3 mJ/cm$^2$. For pump fluences lesser than 3.3
mJ/cm$^2$, data are fitted using Eq. \eqref{Te_DR_R_chirp} by
taking only one coherent phonon (A$_1$). The fits as shown in time
and frequency domain in Fig. \ref{Te_Fig1} are excellent.

The lattice displacement of the coherent phonon modes can
be estimated  for absorbing materials like bismuth from
\cite{Te_DeCamp, Te_Decamp_Thesis}
\begin{equation}\label{Te_U1}
    \textnormal{U}_i^2 \sim \frac{3.8 \times 10^{-3}~B_i~F}{\varrho \nu_i |\varepsilon|}
\left[\frac{(\frac{2\varepsilon_2}{E_{ph}})}{D}\right]
\end{equation}
where U$_i$ is in Angstrom (\AA), $F$ is the pump fluence in
mJ/cm$^2$, $\varrho$ is the density of the material in amu/\AA$^3$
and $\varepsilon$ is the dielectric constant ($\varepsilon =
\varepsilon_1+j\varepsilon_2$), E$_{ph}$ is the energy of the
phonon in eV and D = $\frac{1}{R}\frac{\partial R}{\partial E}$
with E as the photon energy in eV. Now, for Te at 1.57 eV, D =
$\frac{1}{R}\frac{\partial R}{\partial E} \sim 10^{-1} eV^{-1}$,
$\left(\frac{2\varepsilon_2}{E_{ph}}\right) \sim 10^3 eV^{-1}$,
and $\varepsilon$ = 32 + $j$11 ($|\varepsilon|$ = 34) at 300K
\cite{Te_Absorption_PR}. The value of $|\varepsilon|$ at 10K is 27
\cite{Te_Absorption_PR}. Since the exact temperature dependence of
$|\varepsilon|$ is not available, we have used $|\varepsilon| =
34$ at all temperatures. This will introduce a maximum error of
$\sim$ 10$\%$ at 10K. The final expression for the lattice
displacement in this case is
\begin{equation}\label{Te_equation5}
\textnormal{U}_i \sim \sqrt{B_i \frac{38~F}{\varrho \nu_i
|\varepsilon|}}
\end{equation}
This gives a very high value of U$_2$ $\sim$ 0.24 \AA~which is
almost twice of $\sim$ 0.13 $\AA$~estimated in Bi
\cite{Te_M_Hase_PRL_2002}. In terms of the lattice constant
\emph{a} = 4.4561~\AA, U$_2$ $\sim$ 0.05\emph{a} which is
unusually very high. The oscillation amplitude of E$_{TO}^1$ and
E$_{TO}^2$ are, B$_1$= 0.03 and B$_3$= 0.01 respectively
corresponding to lattice displacements of U$_1 \sim$ 0.11~\AA~and
U$_3 \sim$ 0.05~\AA. Thus the huge lattice displacements caused by
the high carrier density excitation may be responsible for
observing the forbidden E$_{TO}$ modes in isotropic reflectivity
detection scheme \cite{Te_M_Hase_PRL_2002}. The possibility of
these E modes due to major crystal orientation in the
polycrystalline sample is perhaps negligible as these modes were
seen only at 3K with the highest pump-fluence. Since E-modes are
absent at all lower pump fluences and temperatures higher than 3K,
we proceed now to discuss only the A$_1$ mode's behavior. From
here, we make a change in notation for the fit parameters
corresponding to phonons as, B$_{ph}$, $\tau_{ph}$, $\nu_{ph}$,
$\beta$ and $\phi$. The fit parameters of the non-oscillatory
electronic part and the oscillatory phononic part are presented in
the following sections.

\subsection{A. Electronic part}
\begin{figure}[htb]
\centering\includegraphics[scale=1.2]{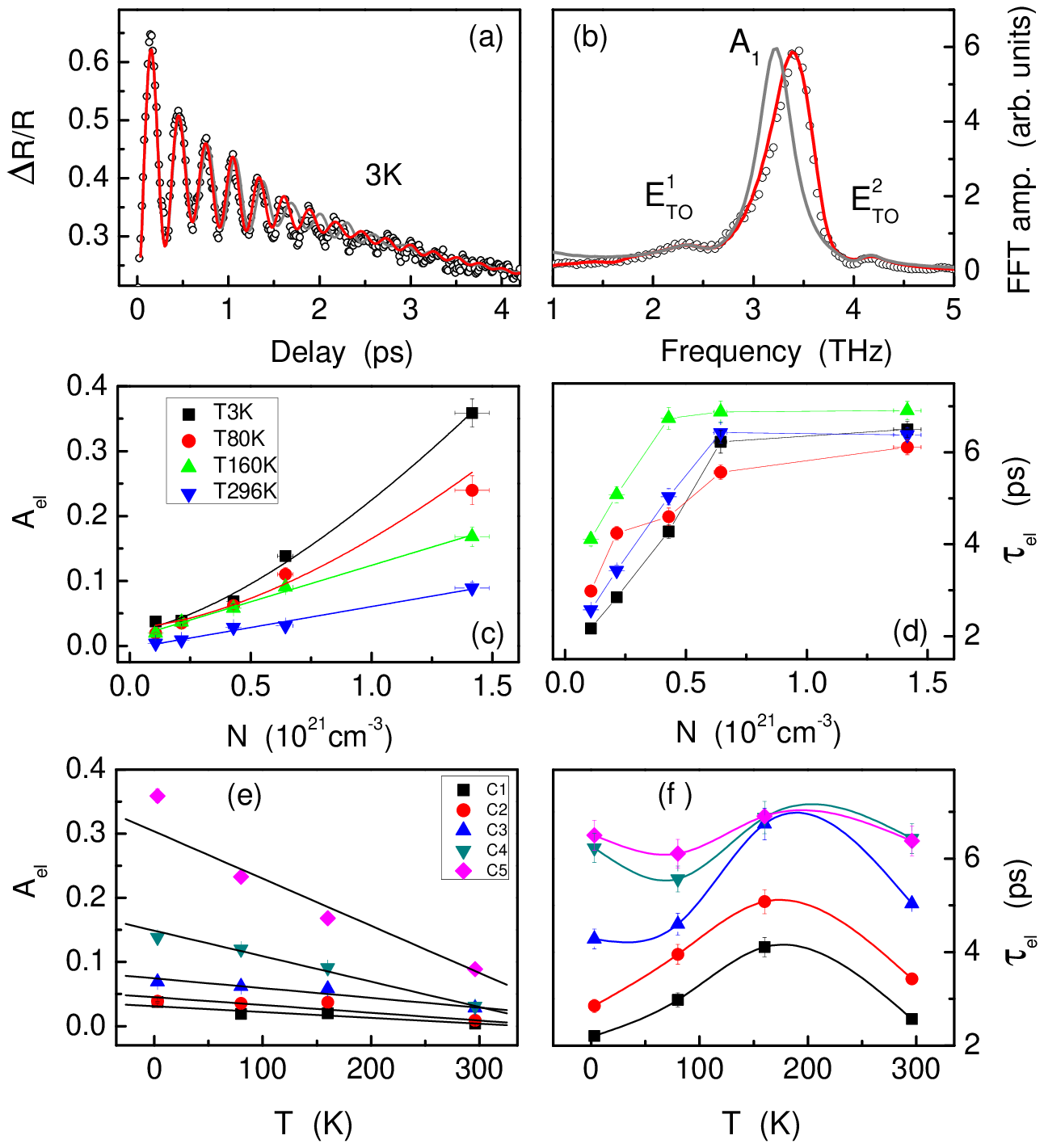} \caption{
}\label{Te_Fig2}
\end{figure}

Fig. \ref{Te_Fig2}(c) and \ref{Te_Fig2}(d) show the dependence of
A$_{el}$ and $\tau_{el}$ on PCD at four different temperatures. It
is seen from Fig. \ref{Te_Fig2}(c) that the strength of the
electronic contribution, A$_{el}$ increases with the pump fluence
at all temperatures. A$_{el}$ is fitted (solid line in the figure)
with A$_{el}\sim$ $N^p$ where $p$ = 1.5 at 3K and 80K; and $p$ =
1.0 at 160K and 296K. The photoexcited carrier relaxation time,
$\tau_{el}$ increases with the PCD ($N$) (Fig. \ref{Te_Fig2}(d))
and saturates at $\sim$ 0.6 $\times$ 10$^{21}$ cm$^{-3}$. Figs.
\ref{Te_Fig2}(e) and \ref{Te_Fig2}(f) display A$_{el}$ and
$\tau_{el}$ \emph{vs} temperature at different PCD (legend C1-C5
corresponds to 0.1, 0.2, 0.4, 0.6 and 1.4 $\times$ 10$^{21}$
cm$^{-3}$). Thick lines in Fig. \ref{Te_Fig2} (e) are the linear
fits (A$_{el}$(T) = A$_{el}(0)$- bT). It is seen from Fig.
\ref{Te_Fig2}(e) that A$_{el}$ increases with lowering of the
sample temperature, reaching a maximum value of 0.36 at 3K (at 1.4
$\times$ 10$^{21}$ cm$^{-3}$), almost 4 times higher than its
value at 296K. The trend is same at other PCDs. We do not yet
understand the temperature and PCD dependences of A$_{el}$. The
dependence of $\tau_{el}$ (Fig. \ref{Te_Fig2}(f)) on temperature
is rather anomalous: the electron relaxation time is seen to
increase with increase in temperature. The solid lines in Fig.
\ref{Te_Fig2} (d) and (f) are drawn as guide to the eye.

\begin{figure}[htb]
\centering\includegraphics[angle=90, scale=0.75]{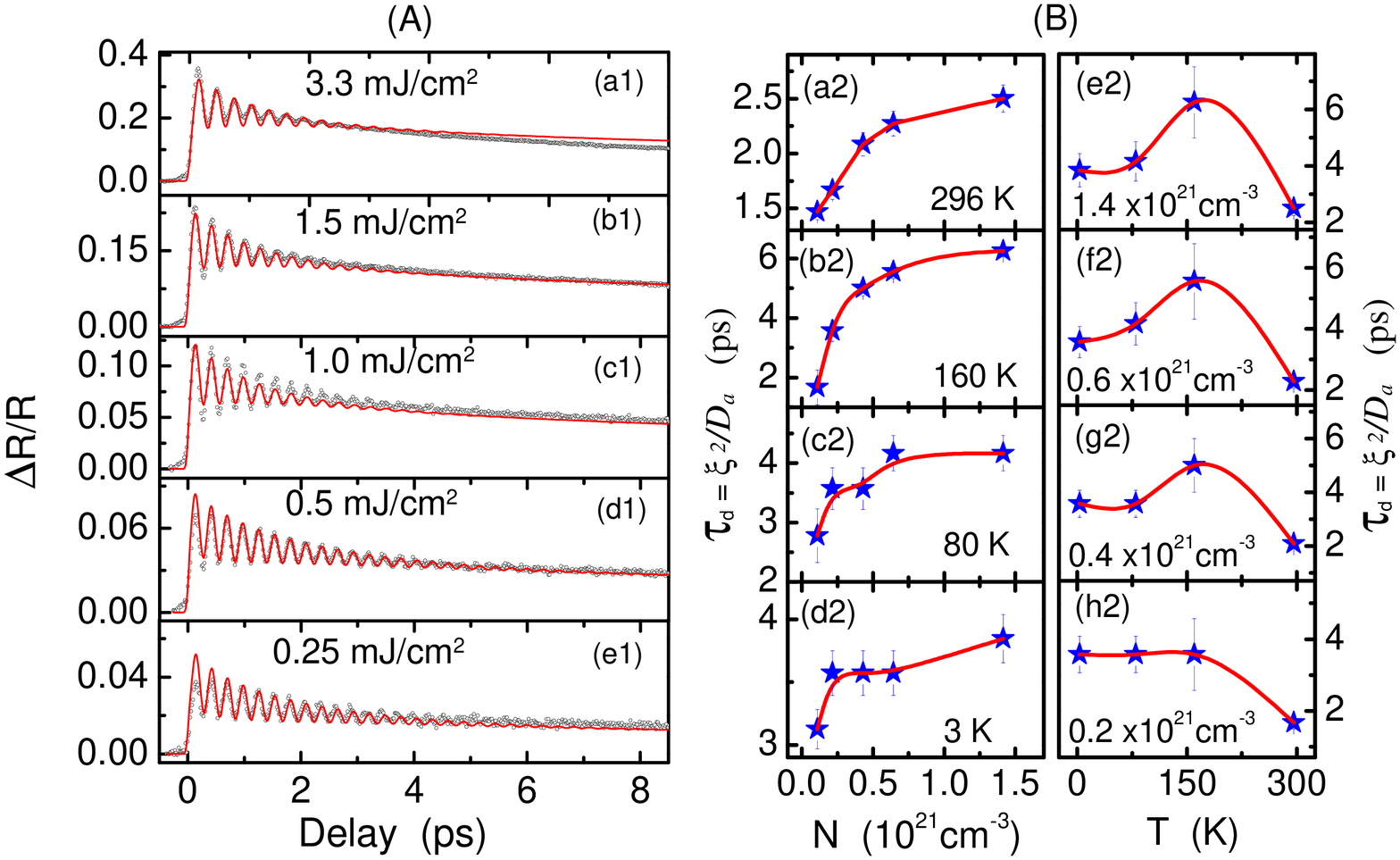}
\caption{}\label{Te_Fig3}
\end{figure}

    We now try to understand carrier dynamics using the
diffusion of photoexcited carriers with time. When a femtosecond
pump pulse irradiates the sample, it creates a carrier density,
$N_0$ on the surface depending on the absorption coefficient and
the fluence. The carrier diffusion leads to a rapid decay of these
carriers across the surface. The diffusion coefficients of
photo-generated electrons and holes can be different leading to a
Dember field \cite{Te_Dember, Te_Kurz_PRL, Te_Kurz_PRB} with a
build up time of 100 fs to 200 fs after the pump excitation. Since
the carrier relaxation times in our experiments are long (2 ps to
6 ps as shown in Fig. \ref{Te_Fig2}(d)), we work with the
assumption of ambipolar diffusion of carriers, described by the
diffusion equation \cite{Te_Xu_APL_V92_2008}
\begin{equation}\label{Te_diffusion}
    \frac{\partial N}{\partial t} = D_a \frac{\partial^2 N}
    {\partial z^2} + \frac{N_0}{\tau_p}e^{(-t/\tau_p)^2} e^{-z/\xi}
\end{equation}
where $\tau_p$ is the pulse width, $D_a$ is the ambipolar
diffusion coefficient and $z$ is the depth inside the
medium. We solved this partial differential equation
numerically for the carrier density N(z,t) with the initial
condition $N(z,0)$= $N_0$ and the boundary conditions,
$D_a\frac{\partial N}{\partial z}$=0 at $z=0$ and and
$N(L,t)=N_0$ e$^{-L/\xi}$. The phonon displacement,
\emph{x}$_{ph}$, given by a damped harmonic oscillator
where the displacement and the angular frequency,
$\Omega_0(N)$ depend on the carrier density $N(z,t)$
 \cite{Te_Tangney_PRB_2002}
 \begin{equation}\label{Te_xph_HO}
    \frac{\partial^2 x_{ph}}{\partial t^2} + \Omega_0(N)^2
    [x_{ph}-x_0(N)]+ \Gamma \frac{\partial x_{ph}}{\partial
    t}= 0
\end{equation}
Here $\Gamma$ is a damping constant for the phonon. In
linear approximation (which is expected to be valid since
the PCD is less than 1$\%$ of the valence electron
density), the carrier density dependence of phonon
displacement, \emph{x}$_0(N)$ and the angular frequency,
$\Omega_0(N)$ are taken as, \emph{x}$_0(N)$=
\emph{x}$_{eq}$ + $(\partial \emph{x}_0/\partial N)N$ and
$\Omega_0(N)= \Omega_{eq} + (\partial\Omega_0/\partial
N)N$; \emph{x}$_{eq}$ and $\Omega_{eq}$ are the lattice
displacement and angular phonon frequency at equilibrium
(at low carrier density levels). We have used
\emph{x}$_{eq}$ = 0.2686 \emph{a} (ref.
\cite{Te_Tangney_PRB_2002}) and the experimentally relevent
values ($\Omega_{eq}$= 23 THz and $\Gamma$ = 0.56 THz and
$\left[\frac{\partial \Omega_0}{\partial N}\right]$ = 1.74
$\times$ 10$^{-21}$ THz-cm$^{3}$ at 160K [see inset of Fig.
\ref{Te_Fig4} (d)]). Eq. \eqref{Te_xph_HO} was solved
numerically using fourth order Runge-Kutta method with the
integration step of 2 fs. The dielectric constant
$\varepsilon = \varepsilon_1 + j \varepsilon_2$ depends on
$x_{ph}$ and $N(z,t)$. Since $\varepsilon_1$ = 32, much
larger than $\varepsilon_2$ (= 11), the changes in
$\varepsilon_1$ will dominate the change in the
reflectivity. Thus,
\begin{eqnarray}
    \label{Te_epsilon_R2}
    \frac{\Delta R}{R} \approx &&
    \varsigma\left[\frac{\Delta\varepsilon_1}{\varepsilon_1}\right]\\
    \label{Te_epsilon_R1}
   \textnormal{with}~~\frac{\Delta\varepsilon_1}{\varepsilon_1} = && \frac{1}{\varepsilon_1}
   \left[\left(\frac{\partial \varepsilon_1}
    {\partial x_{ph}}\right)x_{ph} + \left(\frac{\partial \varepsilon_1}
    {\partial N}\right)N\right]
\end{eqnarray}
where $\frac{\partial \varepsilon_1}{\partial x_{ph}}$,
$\frac{\partial \varepsilon_1}{\partial N}$ and $\varsigma$
are taken as adjustable parameters to fit the
experimentally observed background and initial oscillations
in the reflectivity. As an example, the data at 160K and
the fit using the Eqs.
\eqref{Te_diffusion}-\eqref{Te_epsilon_R1} are shown in
Fig. \ref{Te_Fig3} in panel (A) for different values of
pump fluence. The variation of ambipolar diffusivity, D$_a$
extracted from the fit are used to calculate the diffusion
time $\tau_d~\sim~\xi^2$/$D_a$. The panel (B) of Fig.
\ref{Te_Fig3} displays thus derived (shown by stars)
$\tau_d$ \emph{vs} $N$ (Fig. \ref{Te_Fig3} (a2)-(d2)) and
$\tau_d$ \emph{vs} T (Fig. \ref{Te_Fig3} (e2)-(h2)). Thick
lines are drawn as a guide to the eye and they are not due
to digital smoothing. Comparison of Fig. \ref{Te_Fig3}
(a2)-(d2) with Fig. \ref{Te_Fig2} (d) shows that the values
of $\tau_d$ and $\tau_{el}$ are comparable and their
dependence on $N$ is similar. Similarly, it is seen that
the behavior of $\tau_d$ \emph{vs} T (Fig. \ref{Te_Fig3}
(e2)-(h2)) is compatible with $\tau_e$ \emph{vs} T (Fig.
\ref{Te_Fig2} (f)) except at the lowest pump fluence. From
the above analysis, it is clear that the carrier diffusion
plays a major role in determining the reflectivity changes
in tellurium after the the pump pulse excitation.

\subsection{B. Phononic part}
\begin{figure}[htb]
\centering\includegraphics[scale=1.5]{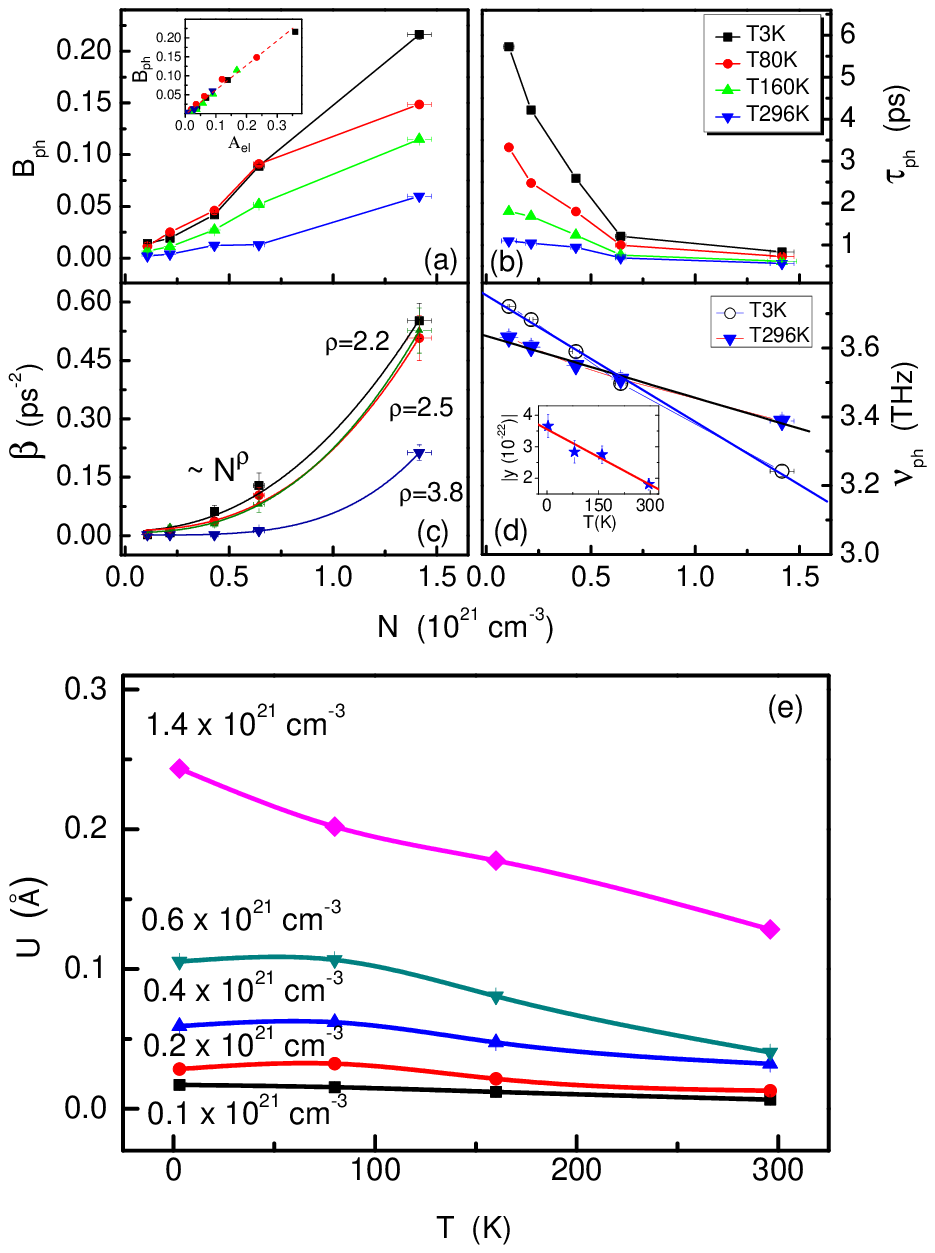}
\caption{}\label{Te_Fig4}
\end{figure}

The dependence of phonon amplitude ($B_{ph}$) on pump
fluence (Fig. \ref{Te_Fig4} (a)) is similar to that of
$A_{el}$ (Fig.\ref{Te_Fig2} (c)). A plot of $B_{ph}$
\emph{vs} $A_{el}$ given in the inset of Fig. \ref{Te_Fig4}
(a) shows a linear dependence between them, implying that
the photoexcited carriers contribute mainly to the
generation of coherent phonons as expected in a DECP
process. The phonon lifetime, $\tau_{ph}$ decreases with
increasing the pump fluence (see Fig. \ref{Te_Fig4} (b))
due to electron-phonon interaction. Thick lines in (a) and
(b) are the guide to the eye. At 3K, the phonon lifetime,
$\tau_{ph}$ is $\sim$ 5.73 ps at the lowest PCD of $\sim$
0.1 $\times$ 10$^{21}$cm$^{-3}$ (corresponding to the
phonon linewidth $\gamma~\equiv \frac{1}{\pi \tau_{ph}}$ =
0.056 THz); this decreases by 87 $\%$ to 0.74 ps at the
highest PCD of 1.4 $\times$ 10$^{21}$cm$^{-3}$. A similar
behavior is seen at all other temperatures.

Next, we turn our attention to the chirp parameter $\beta$
shown in Fig. \ref{Te_Fig4} (c). The chirp parameter
$\beta$ is fitted (solid line) with $\sim$ $N^\rho$ with
$\rho$ = 2.2 (3K), $\rho \sim$ 2.5 (80K and 160K) and
$\rho$ = 3.8 (296K). The phonon chirping increases with
increasing pump fluence; since A$_{el}~\sim N^p$,
$\beta~\sim A_{el}^{\rho/p}$. It can be seen from Fig.
\ref{Te_Fig4} (c) that $\beta$ increases on lowering the
temperature. For example, at highest PCD of 1.4 $\times$
10$^{21}$cm$^{-3}$, $\beta \sim 0.55$ at 3K as compared to
$\beta \sim 0.2$ at 296K.

    We now look at the dependence of the observed phonon frequency
as a function of photo carrier density at two temperatures,
3K and 296K as shown in Fig.\ref{Te_Fig4} (d). Here, the
variation of the frequency is linear with $N$ (linear fits
are shown by solid lines in Fig. \ref{Te_Fig4} (d)), as
taken before in solving the Eq. \eqref{Te_xph_HO}.
$\nu_{ph}$(THz) = $\nu_0$(THz)-y(THz cm$^{3}$)
$N$(cm$^{-3}$) with $\nu_0$ = 3.75 THz and 3.64 THz at 3K
and 296K respectively; and $y \equiv
\frac{1}{2\pi}\left[\frac{\partial \Omega_0}{\partial
N}\right]$ = 3.73 $\times$ 10$^{-22}$ THz-cm$^{3}$ at 3K
and 1.86$\times 10^{-22}$ THz-cm$^{3}$ at 296K. The slope y
determined here corresponds to the change in frequency with
pump fluence as $\sim$ -0.16 THz/mJ-cm$^{-2}$ (3K) and
$\sim$ -0.08 THz/mJ-cm$^{-2}$ (300K). The value at room
temperature agrees well with the theoretically estimated
value \cite{Te_Tangney_PRB_2002} of -0.085 THz/mJ-cm$^{-2}$
at 296K for tellurium. The modulus of the slope, $|y|$ from
this linear fit is plotted as a function of temperature in
the inset of Fig. \ref{Te_Fig4} (d) showing that the
softening of phonons with carrier density decreases
linearly with increasing temperature.

The lattice displacements of A$_1$ mode using Eq.
\eqref{Te_equation5} at various PCD \emph{vs} temperatures
are summarized in Fig. \ref{Te_Fig4} (e) where the solid
lines are connected through the data points as guide to the
eye. Consistent with more photoexcited carriers at 3K (pump
fluence of 3.3 mJ/cm$^2$), U/\emph{a} is $\sim$ 5 $\%$,
which gradually decreases to 2.5 $\%$ at 296K, i.e a
decrease by a factor of 2. The trend is similar for other
pump fluences (see Fig. \ref{Te_Fig4} (e)).

\subsection{C. Low pump fluence experiments}
\begin{figure}[htb]
\centering\includegraphics[scale=1.2]{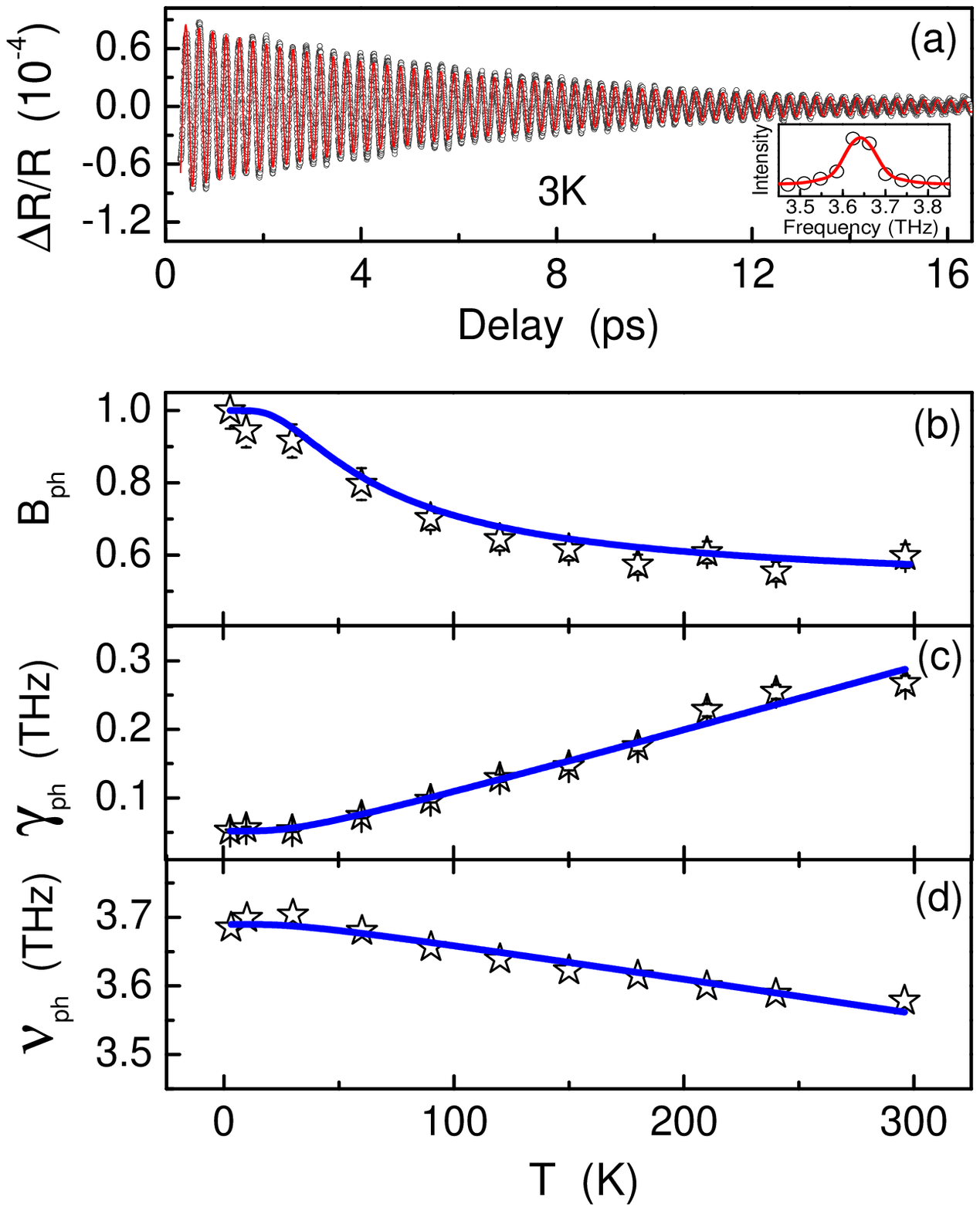}
\caption{}\label{Te_Fig5}
\end{figure}

To separate the effects of electron-phonon and
phonon-phonon anharmonic interactions in the frequency and
lifetime of coherent phonons presented above, we have
performed the degenerate pump-probe reflectivity
experiments with very low pump fluence of 14 $\mu$J/cm$^2$
(N$_0~\sim~6~\times$ 10$^{18}$ cm$^{-3}$). The varying
non-oscillatory electronic background in the data is
removed using FFT digital filter smoothing to extract the
coherent phonon oscillations. Thus obtained data was fitted
using Eq. (\ref{Te_DR_by_R}) as the chirping is absent. The
typical reflectivity data after FFT filtering at 3K along
with the fit are shown in Fig \ref{Te_Fig5} (a). The fitted
parameters- the coherent phonon amplitude B$_{ph}$, damping
parameter $\gamma_{ph}~(\equiv~\frac{1}{\pi\tau_{ph}})$ and
the phonon frequency $\nu_{ph}$ are plotted as a function
of temperature in Fig \ref{Te_Fig5} (b), (c) and (d). It
has been shown earlier \cite{Te_Misochko_JPCM_2006} that
the coherent phonon amplitude, B$_{ph}$ behaves quite
similar to Raman peak intensity as I$_p$ $\sim
\frac{[n(\nu_{ph})+1]}{[2 n(\nu_{ph}/2)+1]}$, where
$n(\nu_{ph}$) is the Bose–Einstein factor. The fit (solid
line) using this expression is shown along with the data
for B$_{ph}$ in Fig. \ref{Te_Fig5}(b) where, B$_{ph}$ is
normalized with respect to its value at 3K. $\gamma_{ph}$
and $\nu$ are fitted (solid line in Fig. \ref{Te_Fig5} (c)
and (d)) with the following well known functions
\cite{Te_anharmonic, Te_Jose_Cardona_PRB_1984} based on
cubic anharmonicity where the phonon of frequency $\nu$
decays into two phonons of equal frequency: $\gamma_{ph}(T)
= \gamma_0 + \Delta\gamma_{anh}(T)$ with
$\Delta\gamma_{anh}(T) = C[1 + 2n(\nu_0/2)]$ and
$\nu_{ph}(T)= \nu_0 + \Delta\nu_{anh}(T)$ with
$\Delta\nu_{anh}(T) = A [1 + 2n(\nu_0/2)]$ where $\nu_0$,
A, C and $\gamma_0$ are the fitting parameters (A and C are
the measures of third order cubic anharmonicity). The
parameters obtained from fitting are $\nu_0$ = 3.70 $\pm$
0.01 THz, A = -0.022 $\pm$ 0.002 THz, $\gamma_0$ = 0.010
$\pm$ 0.003 THz and C = 0.042 $\pm$ 0.003 THz. Fig.
\ref{Te_Fig5}(b), (c) and (d)  show that the fits to the
anharmonic contributions are excellent.

\subsection{D. Electron-phonon coupling}
In doped semiconductors, the electron-phonon interactions
contribute to the linewidth ($\gamma_i$) of the i$^{th}$
phonon of frequency $\nu_i$ and degeneracy g$_i$ which is
related to the dimensionless electron-phonon coupling
constant $\lambda_{EP}$ and density of states at the Fermi
level (DOS($\epsilon_F$)) by the Allen's formula
\cite{Te_Allen}: $\gamma_i = (2\pi/ g_i)\lambda_{EP}\nu_i^2
DOS(E_{F})$. The question arises what happens to phonons
due to the photoexcited carriers under femtosecond pulse
excitation. We address this question and extract the
contribution of electron-phonon interaction to the phonon
frequency and linewidth as described below.

\begin{figure}[htb]
\centering\includegraphics[scale=1.2]{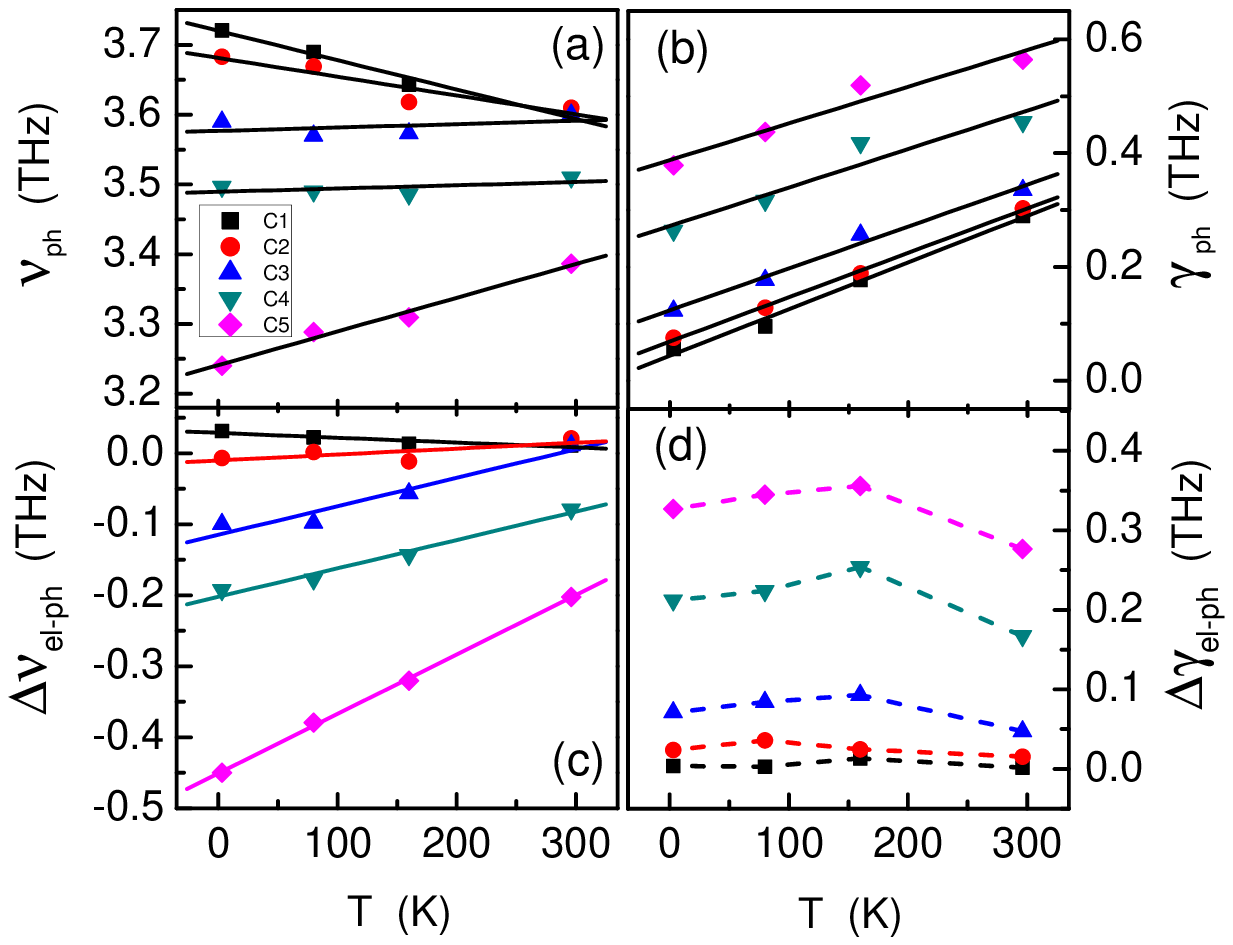}
\caption{}\label{Te_Fig6}
\end{figure}

The temperature dependance of $\nu_{ph}$ and $\gamma_{ph}$ along
with their linear fits at various pump fluences are drawn in Fig.
\ref{Te_Fig6} (a) and (b)(legends C1-C5 correspond to 0.1, 0.2,
0.4, 0.6 and 1.4 $\times$ 10$^{21}$ cm$^{-3}$). We note that the
effect of the carrier diffusion is explicitly taken into account
while extracting the phonon parameters, $\nu_{ph}$ and
$\gamma_{ph}$ (shown in Fig. \ref{Te_Fig6} (a) and \ref{Te_Fig6}
(b)) by fitting the total $\frac {\Delta R} {R}$ with Eq.
\ref{Te_DR_R_chirp}. It can be seen that the normal anharmonic
behavior of phonons (i.e, frequency increasing with lowering
temperatures) at the lowest pump fluence of 0.1 $\times$ 10$^{21}$
cm$^{-3}$ changes gradually to an anomalous behavior (frequency
decreasing with lowering temperatures) as the pump fluence is
increased, suggesting a strong electron-phonon coupling at higher
carrier densities. No such anomalous behavior is noticed in the
phonon damping term (Fig. \ref{Te_Fig6} (b)). The phonon frequency
$\nu_{ph}(T)$ and damping $\gamma_{ph}(T)$ thus deduced have
contributions both from electron-phonon and phonon-phonon
anharmonic interactions. The temperature variation of frequency
and linewidth of a phonon mode can be thus written as
\cite{Te_anharmonic, Te_EPC_Ferrari}
\begin{eqnarray}
    \nu_{ph}(T) & = & \nu_{0} + \Delta\nu_{anh}(T) +
    \Delta\nu_{el-ph}(T)\label{Te_anhar1}\\
    \gamma_{ph}(T) & = & \gamma_0 + \Delta\gamma_{anh}(T) +
    \Delta\gamma_{el-ph}(T)\label{Te_anhar2}
\end{eqnarray}
where $\nu_{el-ph}$ and $\gamma_{el-ph}$ are due to
electron-phonon interaction.

    For the low-pump fluence case of 14 $\mu$ J/cm$^2$ (data shown in
Figs. \ref{Te_Fig5} (c) and (d)), the contribution of
electron-phonon coupling is negligible and the temperature
dependence arises predominantly from anharmonic interactions.
Thus, by subtracting the values of $\nu_{ph}(T)$ and
$\gamma_{ph}(T)$ shown in Figs. \ref{Te_Fig5} (d) and
\ref{Te_Fig5} (c) from the corresponding values shown in Figs.
\ref{Te_Fig6} (a) and \ref{Te_Fig6} (b), we obtain only the
electron-phonon contribution to the phonon parameters (shown in
Figs. \ref{Te_Fig6} (c) and \ref{Te_Fig6} (d)) at different
carrier density levels. The solid lines in Fig. \ref{Te_Fig6} (c)
are fit to $\Delta\nu_{el-ph}(T)~=~\Delta\nu_{el-ph}(0)+A(N)T,$
where the slope A is a function of photoexcited carrier density N.
It can be seen that both $\Delta\nu_{el-ph}$ and
$\Delta\gamma_{el-ph}$ strongly depend on the carrier density N.
On the other hand, $\Delta\gamma_{el-ph}$ is weakly dependent on
T.  More theoretical understanding is needed to explain the trends
seen in Fig. \ref{Te_Fig6} (c) and (d).

\subsection{V. CONCLUSIONS}
Time-resolved reflectivity measurements of tellurium have
been performed over a wide range of temperatures and
photoexcited carrier densities. The relaxation time
associated with the carrier diffusion increases with the
pump fluence which is quantitatively understood based on a
diffusion model. The temperature dependence of carrier
relaxation time, $\tau_{el}$, is observed to be anomalous.
The oscillatory part of the time-resolved differential
reflectivity shows highly chirped A$_1$ phonons excited at
high carrier densities and the chirping is found to be
largest at the lowest temperature where the carrier
diffusion time and the strength of photoexcited carrier
density are maximum. Apart from the chirping seen in
phonons, the lattice displacements associated with them is
estimated to be very high (e.g $\sim$ 0.24 \AA~ at 3K).
Another important result is the observation of increased
softening of $A_1$ mode with the pump fluence at 3K (-0.16
THz/mJ-cm$^{-2}$) compared to 296K (-0.08 THz/mJ-cm$^{-2}$)
consistent with the increased photoexcited carriers at low
temperatures. Using low pump fluence transient reflectivity
experiments, the effects of phonon anharmonicity and the
contribution of electron-phonon interaction to the phonon
frequencies and linewidths at high pump fluence were
separated. The temperature and photoexcited carrier density
dependences of electron-phonon contribution to the mode
frequency and linewidth needs to be understood
theoretically.

\subsection{ACKNOWLEDGEMENTS}
AKS acknowledges the financial support from Department of Science
and Technology of India. SK acknowledges University Grants
Commission, India for senior research fellowship.

\newpage
FIGURE CAPTIONS:

Figure 1:~(a) Normalized time-resolved reflectivity data
(open circles) at T=3K at various pump fluences along with
the fits (line) according to Eq. (\ref{Te_DR_R_chirp}). (b)
Corresponding FFT spectra of the oscillatory part of the
data and fit. The vertical dashed line in (b) is the guide
to the eye to appreciate the red shift of the phonon peak
frequency.

Figure 2:~(a) Normalized time-resolved reflectivity change
($\frac{\Delta R}{R}$) at T=3K at maximum pump fluence of
3.3 mJ/cm$^{2}$ along with the fit.(b) The FFT spectra of
the oscillatory part of time domain part. Open circles
represent the experimental data and thick line is the fit
according to Eq. (\ref{Te_DR_R_chirp}) with $\beta = 0.55$
and the grey line is with $\beta = 0$. (c) Carrier density
dependence of strength of photoexcited carriers (A$_{el}$)
and (d) photoexcited carrier relaxation time ($\tau_{el}$)
at different temperatures (T=3K (filled squares), 80K
(filled circles), 160K (filled triangles), 296K (filled
inverted triangles)). The fit, A$_{el}~\sim N ^p$ is shown
in (c) with p =1.5~(3K), 1.6~(80K), 1 (160K and 296K). The
lines in (d) are guide to the eye. (e) The temperature
dependence of A$_{el}$ and (f) $\tau_{el}$ at given carrier
densities (C1 = 0.1$\times$10$^{21}$cm$^{-3}$, C2 =
0.2$\times$10$^{21}$cm$^{-3}$, C3 =
0.4$\times$10$^{21}$cm$^{-3}$, C4 =
0.6$\times$10$^{21}$cm$^{-3}$, C5 =
1.4$\times$10$^{21}$cm$^{-3}$). The lines in (e) are the
linear fits and in (f) are guide to the eye.

Figure 3:~(A) Time domain data at 160K (open circles) and
the corresponding fit (line) using the diffusion model (see
text) at different fluences (a1)-(e1). (B) The diffusion
time (filled stars), $\tau_d$ = $\xi^2$/$D_a$ versus N
(a2)-(d2) and T (e2)-(h2) (see text). Here $D_a$ is the
carrier diffusion coefficient and the solid lines are drawn
as guide to the eye.

Figure 4:~PCD dependance of phonon fit parameters: (a)
phonon amplitude ($B_{ph}$), (b) phonon life time
($\tau_{ph}$), (c) phonon chirping parameter ($\beta$)
along with a fit $\sim N^\rho$ (2.2 $\leq\rho\geq$ 3.8),
and (d) the phonon frequency, $\nu_{ph}$ at two
temperatures-296K and 3K. The inset of (d) shows the
dependance of the slope (filled stars),
$y~(=\frac{1}{2\pi}\frac{\partial \Omega_0}{\partial N}$
THz cm$^{3}$) of the linear fits to the data as function of
temperature (solid line is the linear fit). (e) The
variation of the lattice displacement, $U$ as a function of
the sample temperature at different PCDs. The data in (a),
(b) and (e) are connected by solid lines as a guide to the
eye.

Figure 5:~(a) Time domain data at lowest pump fluence of 14
$\mu$J/cm$^2$ at 3K (open circles) along with fit to Eq.
\eqref{Te_DR_by_R} (solid line). Inset is the FFT of the
data and the fit. (a) Phonon amplitude, B$_{ph}$ vs T, (c)
phonon damping parameter, $\gamma_{ph}$ vs T (d) the phonon
frequency $\nu_{ph}$ vs T. The solid lines are based on
cubic anharmonic fit (see text) and the open stars are the
data.

Figure 6:~ Temperature dependence of (a) $\nu_{ph}$, (b)
$\gamma_{ph}$,  (c) $\Delta\nu_{el-ph}$ and (d)
$\Delta\gamma_{el-ph}$ at different carrier density levels.
Legends C1-C5 corresponds to 0.1, 0.2, 0.4, 0.6 and 1.4 $\times$
10$^{21}$ cm$^{-3}$. The solid lines in (a), (b) and (c) are the
linear fits. The dashed lines in (d) are the guide to the eye.

\newpage

\begin{thebibliography}{10}
\providecommand*{\bibinfo}[2]{#2}
\providecommand*{\eprint}[1]{#1}
\providecommand*{\url}[1]{#1}

\bibitem{Te_A_Othonos_JAP_Review_1998}
A. Othonos, J. Appl. Phys. \textbf{83}, 1789 (1998).

\bibitem{Te_Knox_PRL}
W. H. Knox, C. Hirlimann, D. A. B. Miller, J. Shah, D. S.
Chemla, and C. V. Shank, Phys. Rev. Lett. \textbf{56}, 1191
(1986).

\bibitem{Te_Oudar_PRL}
J. L. Oudar, D. Hulin, A. Migus, A. Antonetti, and F.
Alexandre, Phys. Rev. Lett. \textbf{55}, 2074 (1985).

\bibitem{Te_J_Shaw}
J. Shah, Ultrafast Spectroscopy of Semiconductors and
Semiconductor Nanostructures (Springer Verlag, Berlin,
1996).



\bibitem{Te_Nelson}
L. Dhar, J. A. Rogers and K. A. Nelson, Chem. Rev.
\textbf{94}, 157 (1994).

\bibitem{Te_DECP_PRB_1992}
H. J. Zeiger, J. Vidal, T. K. Cheng, E. P. Ippen, G.
Dresselhaus, M. S. Dresselhaus, Phys. Rev. B. \textbf{45}
768 (1992).

\bibitem{Te_Cheng_APL}
T. K. Cheng, M. S. Dresselhause, and E. P. Ippen, Appl.
Phys. Lett. \textbf{62}, 1901 (1993).

\bibitem{Te_Kuznetsov}
A. V. Kuznetsov and C. J. Stanton, Phys. Rev. Lett.
\textbf{73}, 3243 (1994).


\bibitem{Te_Garret_PRL_1996}
G. A. Garrett, T. F. Albrecht, J. F. Whitaker, and R. Merlin,
Phys. Rev. Lett. \textbf{77}, 3661 (1996).

\bibitem{Te_Dember}
H. Dember, Phys. Z. \textbf{32}, 554 (1931).

\bibitem{Te_Kurz_PRL}
T. Dekorsy, H. Auer, C. Waschke, H. J. Bakker, H. G. Roskos, and
H. Kurz, Phys. Rev. Lett. \textbf{74}, 738 (1995).

\bibitem{Te_Kurz_PRB}
T. Dekorsy, H. Auer, H. J. Bakker, H. G. Roskos, and H. Kurz,
Phys. Rev. B \textbf{53}, 4005 (1996).

\bibitem{Te_Hunsche}
S. Hunsche, K. Wienecke, T. Dekorsy, and H. Kurz, Phys.
Rev. Lett. \textbf{75}, 1815 (1995).

\bibitem{Te_Hunsche_APA}
S. Hunsche,K. Wienecke, and H. Kurz, Appl. Phys. A,
\textbf{62}, 499 (1996).

\bibitem{Te_Tangney_PRB_2002}
P. Tangney, S. Fahy, Phys. Rev. B \textbf{65}, 054302
(2002).


\bibitem{Te_DeCamp} M. F. DeCamp, D. A. Reis, P. H. Bucksbaum, and
R. Merlin, Phys. Rev. B \textbf{64}, 092301 (2001).

\bibitem{Te_M_Hase_PRL_2002} M. Hase , M. Kitajima, S. I. Nakashima, and
K. Mizoguchi, Phys. Rev. Lett. \textbf{88}, 067401 (2002).

\bibitem{Te_Misochko_PRL_2004} O. V. Misochko , M. Hase, K. Ishioka,
and M. Kitajima, Phys. Rev. Lett. \textbf{92}, 197401
(2004).

\bibitem{Te_Murray} E. D. Murray, D. M. Fritz, J. K. Wahlstrand,
S. Fahy, and D. A. Reis, Phys. Rev. B \textbf{72}, 060301
(2005).

\bibitem{Te_Stampfi_PRB_electron_Softening}
P. Stampfli and K. H. Bennemann, Phys. Rev. B \textbf{42},
7163 (1990); \textbf{46}, 10686 (1992); \textbf{49}, 7299
(1994).

\bibitem{Te_pine_raman}
A. S. Pine and G. Dresselhaus, Phys. Rev. B \textbf{4}, 356
(1971).

\bibitem{Te_Absorption_PR} S. Tutihasi, G. G. Roberts, R. C.
Keezer, and R. E. Drews, Phys. Rev.  \textbf{177}, 1143
(1969).

\bibitem{Te_Shaker}
G.C. Cho, W. K$\ddot{u}$tt, and H. Kurz, Phys. Rev. Lett.
\textbf{65}, 764(1990).

\bibitem{Te_Decamp_Thesis}
M. F. Decamp, Ph.D Thesis, The University of Michigan, Ann
Arbor (2002).

\bibitem{Dekorsy_PRL_V74_738}
T. Dekorsy, H. Auer, C. Waschke, H.J. Bakker, H.G. Roskos,
H. Kurz, V. Wagner, and P. Grosse, Phys. Rev. Lett.
\textbf{74}, 738 (1995).

\bibitem{Te_Xu_APL_V92_2008}
A. Q. Wu, X. Xu, and R. Venkatasubramanian, Appl. Phys.
Lett. \textbf{92}, 011108 (2008).

%

\bibitem{Te_Misochko_JPCM_2006}
O. V. Misochko, K. Ishioka, M. Hase, and M. Kitajima, J.
Phys.: Condens. Matter \textbf{18}, 10571 (2006).

\bibitem{Te_anharmonic}
M. Balkanski, R. F. Wallis, E. Haro, Phys. Rev. B
\textbf{28}, 1928 (1983).

\bibitem{Te_Jose_Cardona_PRB_1984}
J. Men$\acute{e}$ndez and M. Cardona, Phys. Rev. B
\textbf{29}, 2051 (1984).

\bibitem{Te_Allen}
P. B. Allen, Phys. Rev. B \textbf{6}, 2577 (1972).

\bibitem{Te_EPC_Ferrari}
M. Lazzeri, S. Piscanec, F. Mauri, A. C. Ferrari, and J.
Robertson, Phys. Rev. B \textbf{73}, 155426 (2006).

\end{thebibliography}
\end{document}